\documentclass[conference]{IEEEtran}
\usepackage{cite}
\usepackage[dvips]{graphicx}
\graphicspath{{../eps/}}
\DeclareGraphicsExtensions{.eps}
\usepackage[cmex10]{amsmath}
\usepackage{listings}
\usepackage{color}
\usepackage{caption3}
\DeclareCaptionFormat{listings} {\parbox{\textwidth}{\vspace{1cm}#1#2#3}}
\usepackage{algorithmic}
\usepackage{algorithm}
\usepackage{array}
\usepackage[caption=false,font=footnotesize]{subfig}
\usepackage{multirow}
\usepackage{textcomp}
\begin{document}

\title{Optimising Performance Through Unbalanced Decompositions}

\author{\IEEEauthorblockN{Adrian Jackson\IEEEauthorrefmark{1},
Joachim Hein\IEEEauthorrefmark{1}\IEEEauthorrefmark{2}, Colin Roach\IEEEauthorrefmark{3}}
\IEEEauthorblockA{\IEEEauthorrefmark{1}EPCC, 
The University of Edinburgh, Kings Buildings, Mayfield Road, Edinburgh, EH9 3JZ, UK}
\IEEEauthorblockA{\IEEEauthorrefmark{2}Lunds Universitet, Box 118, 221 00 Lund, Sweden}
\IEEEauthorblockA{\IEEEauthorrefmark{3}EURATOM/CCFE Fusion Association, Culham Science Centre, Abingdon, Oxon,
OX14 3DB, UK.}}
\maketitle

\begin{abstract}
GS2 is an initial value gyrokinetic simulation code developed to study low-frequency turbulence in magnetized plasma. 
It is parallelised using MPI with the simulation domain decomposed across tasks. The optimal domain decomposition 
is non-trivial, and complicated by the differing requirements of the linear and non-linear parts of the calculations. 

GS2 users currently choose a data layout, and are guided towards process counts that are efficient for linear 
calculations. These choices can, however, lead to data decompositions that are relatively inefficient for the non-linear calculations. 
We have analysed the impact of the data decompositions on the non-linear calculation performance and associated communications. 
This has helped us to optimise the decomposition algorithm by using unbalanced data layouts for the non-linear calculations whilst maintaining 
the existing decompositions for the linear calculations. This has completely eliminated 
communications for parts of the non-linear simulation and improved performance by up to $15\%$ for a representative simulation.

\end{abstract}

\IEEEpeerreviewmaketitle

\section{Introduction}
Many scientific simulation techniques, and therefore scientific simulation codes, use spectral methods to solve partial 
differential equations (PDEs) efficiently.  Whilst there are a variety of different techniques to achieve this, the general 
approach is to use a discrete and finite set of basis functions.  This finite set of basis functions allows the PDE to be 
treated numerically and the solution of the PDE is approximated as a linear combination of the basis functions.  Using basis functions with 
global support is commonly referred to as spectral methods.  With periodic boundary conditions, trigonometric functions 
are a popular choice for the basis.  To transfer the numerical solution from the spectral space to the position space one 
typically deploys fast Fourier transformations (FFT).

However, there are a range of algorithms or simulation codes that cannot or do not rely solely on spectral techniques to solve their 
problems.  For instance, operator splitting can be used to separate a system of PDEs into simpler subproblems that can be 
solved separately, using different algorithms.  In this paper we are proposing an optimised data decomposition strategy for 
just one such problem, where the GS2 code (described in Section \ref{sec:GS2} uses operator splitting to separate linear and 
non-linear calculations, which are performed separately.  Maximum efficiency is achieved if the linear part of the calculation is 
performed in Fourier space, and the non-linear part of the calculation is performed in position space.

Whilst, in theory, this should enable the overall solution to be calculated efficiently, the use of two different data spaces (Fourier and
position space) complicates the way data is stored, especially as GS2 is a parallel code with its domain decomposed across processes. The GS2
domain decomposition is usually, and most easily, optimised for the linear calculations. The particular choice of domain decomposition can 
have a significant impact on the communication overhead that is required to transform the data between Fourier and position space for the 
evaluation of the non-linear terms.

In this paper we will briefly describe GS2 in Section \ref{sec:GS2}, then go on in section \ref{sec:PerfAnal} to analyse the 
cost of converting between Fourier and position space in GS2 and in particular the impact of the chosen data layout 
and number of MPI tasks.  Then we present an optimised data decomposition 
to reduce the costs of the transform between Fourier and position space in Section \ref{sec:Unbalanced}, and go on to 
discuss the performance of our new functionality in Section \ref{sec:UnbalancedPerfAnal}.  Finally we summarise our 
work in Section \ref{sec:Conclusions}.

\section{GS2}
\label{sec:GS2}
GS2 \cite{Kotschenreuther_1995}\cite{Dorland_Jenko_Kotschenreuther_Rogers_2000} is an initial value simulation code developed 
to study low-frequency turbulence in magnetized plasma. 
GS2 solves the gyrokinetic equations for perturbed distribution functions together with Maxwell's equations 
for the turbulent electric and magnetic fields within a plasma.  It is typically used to assess the microstability 
of plasmas produced in the laboratory and to calculate key properties of the turbulence which results from instabilities. 
It is also used to simulate turbulence in plasmas which occur in nature, such as in astrophysical and magnetospheric systems.

Gyrokinetic simulations solve the time evolution of the distribution functions of each charged particle species in 
the plasma, taking into account the charged particle motion in self-consistent magnetic and electric fields. These 
calculations are undertaken in a five dimensional data space, with three dimensions accounting for the physical 
location of particles and two dimensions accounting for the velocity of particles. More complete six dimensional kinetic plasma 
calculations (with three velocity space dimensions) are extremely time consuming (and therefore computational very costly) not 
least because of the very rapid gyration of particles under the strong magnetic fields.  Therefore, gyrokinetic simulations 
simplify this six dimensional data space by averaging over the rapid gyration of particles around magnetic field lines and 
therefore reducing the six dimensional problem to a five dimensions (where velocity is represented by energy and 
pitch angle, and the gyrophase angle is averaged out). While the fast gyration of particles is not calculated in full, it 
is included in a gyro-averaged sense allowing lower frequency perturbations to be simulated faithfully at lower computational cost.

GS2 can be used in a number of different ways, including linear microstability simulations, where growth rates are 
calculated on a wavenumber-by-wavenumber basis with an implicit initial-value algorithm in the ballooning (or `flux-tube') 
limit. Linear and quasilinear properties of the fastest growing (or least damped) eigenmode at a given wavenumber may be 
calculated independently (and therefore reasonably quickly).  It can also undertake non-linear gyrokinetic simulations of 
fully developed turbulence.  All plasma species (electrons and various ion species) can be treated on an equal, gyrokinetic footing. 
Non-linear simulations provide fluctuation spectra, anomalous (turbulent) heating rates, and species-by-species anomalous 
transport coefficients for particles, momentum, and energy.  However, full non-linear simulations are very computationally 
intensive so generally require parallel computing to complete in a manageable time.

GS2 is a fully parallelised, open source, code written in Fortran90.   The parallelisation is implemented with MPI, with 
the work of a simulation in GS2 being split up (decomposed) by assigning different parts of the domain to different 
processes.

\subsection{Data Layouts and Decomposition}
\label{sec:sub:layoutanddecomp}
GS2 parallelises the array storing the perturbed distribution functions for all the plasma species in 5 different indices, 
denoted as follows by the characters x,y,l,e and s: 
\begin{itemize}
\item {\tt x}: Fourier wavenumber in the X direction in space
\item {\tt y}: Fourier wavenumber in the Y direction in space
\item {\tt l}: Pitch angle
\item {\tt e}: Energy
\item {\tt s}: Number of particle species
\end{itemize}
GS2 supports six different data layouts. These layouts are: {\tt xyles}, {\tt yxles}, {\tt lyxes}, {\tt yxels}, {\tt lxyes}, {\tt lexys}. The layout can be chosen at run 
time by the user (through the input parameter file). 

The layout controls how the data domain in GS2 is distributed across processes, by specifying the 
order in which individual dimensions in the data domain are distributed (split up).  For instance, the 
{\tt xyles} layout will decompose {\tt s} first and {\tt x} last (depending on the number of processes used), whereas the 
{\tt lexys} layout will decompose {\tt s} first and {\tt l} last.

Linear GS2 calculations are performed in Fourier space (k-space) as this is computationally simpler and more efficient.  
The non-linear terms, however, are more efficiently calculated in position space.  Therefore, when GS2 undertakes a 
non-linear simulation, it calculates in the linear advance in k-space, and the non-linear terms in position space. While the 
majority of the simulation is in k-space, at each timestep data must be transformed into position space to evaluate the 
non-linear term, and back into k-space.  Whilst users should not generally have to concern themselves with these implementation 
details, this use of both k- and position space can impact on performance, depending on the number of processes used in the 
parallel computation.

In the linear part of calculations the GS2 data space is parallelised in k-space, using the GS2 {\tt g\_lo} layout. The non-linear 
parts of calculations, which require data in position space, use two data distribution layouts (along with Fast Fourier 
Transforms (FFTs)) to transform data between k-space and position space, and these are called {\tt xxf\_lo} and {\tt yxf\_lo}.  These are required 
for the two stages of the FFT which takes data from the k-space layout ({\tt g\_lo}) to the position space layout 
{\tt yxf\_lo}, via the intermediate {\tt xxf\_lo} layout. This process must also be reversed back to k-space once the non-linear calculations 
have been undertaken.   

This process of calculating a 2D FFT uses both MPI communications and local data copying followed by an FFT to transform 
the data from {\tt g\_lo} space to {\tt xxf\_lo} space.  Then the conversion from {\tt xxf\_lo} space to {\tt yxf\_lo} space, again using 
MPI communications and local data copying, performs the data transposition required for the second FFT. 
Finally, this data is then used by an FFT function to convert the data into real space.  The reverse operations are performed to convert 
the data back to k-space once the non-linear terms have been computed.

Practically the GS2 data space is stored in a single array, but conceptually it is 7 dimensional data object.  
Five of the dimensions are the previously discussed {\tt x,y,l,e,s}, the other 2 are {\tt ig} (the index corresponding 
to the spatial direction parallel to the magnetic field) and {\tt isgn}
({\tt isgn} corresponds to the direction of particle motion in the direction parallel to the magnetic field, $\mathbf{b}$; for GS2 {\tt isgn} 
=1 represents particles moving parallel to $\mathbf{b}$, and {\tt isgn}=2 represents particles moving antiparallel to $\mathbf{b}$).  For the {\tt g\_lo}
data space, the indices {\tt ig} and {\tt isgn} are kept local to each process and {\tt x,y,l,e,s} are decomposed 
(with the order of the decomposition depending on the chosen layout).  
For the {\tt xxf\_lo} data space {\tt x} is kept local and {\tt y,g,isgn,l,e,s} are decomposed (with the order of the {\tt l,e,s}
decomposition depending on the chosen layout).  For the {\tt yxf\_lo} data space {\tt y} is kept local and {\tt x,ig,isgn,l,e,s}
are decomposed (again with the order of {\tt l,e,s} decomposition determined by the chosen layout).  

GS2 also uses a dealiasing algorithm to filter out high wavenumbers in {\tt X} and {\ Y} which is a standard technique to avoid 
non-linear numerical instabilities in spectral codes.  The {\tt g\_lo} data layout is the filtered data, which has lower
  ranges of indices in x and y than in the {\tt xxf\_lo} and {\tt yxf\_lo} layouts.  This means that after the 
{\tt yxf\_lo} stage the amount of data is larger than at the {\tt g\_lo} stage (approximately $2\frac{1}{4}$ times larger).

For optimal performance of the transformation of data between the linear and non-linear parts of the calculation, it is critical to 
minimise the amount of data that is sent using MPI communications, and
therefore to maximise the amount of data that is moved using 
local data copying.  Consider the case where only one process is used to run the program.  In this case all the data 
resides on a single process so the transformation between k-space and position space only requires local data copies, moving 
data around in arrays to the format required for the calculation.  As more processes share the data, the transformation between k- and position
  spaces is increasingly likely to need MPI communication of data between processes.

For any given input file (which specifies the layout used) users generally run a tool/program called {\tt ingen} that provides a list 
of recommended process counts (or "sweet spots") for the simulation to be run on.  These recommendations are calculated from the 
input parameters of the simulation, aiming to split the data domain as evenly as possible to achieve good "load-balancing".  
The primary list of process counts suggested are based on the main k-space data layout where the linear calculations are undertaken ({\tt g\_lo}).  
{\tt ingen} also provides lists of process counts that are suitable for the non-linear non-linear layouts {\tt xxf\_lo} and {\tt yxf\_lo}, which may differ from
the optimal process counts for {\tt g\_lo}.

For optimal performance with GS2, choosing a process count that is common to all three of {\tt ingen}'s lists of suggested process counts 
would be beneficial.  This may not always be possible, especially at larger process counts, so in this scenario 
users tend to choose a process count that is optimal for the linear calculations ({\tt g\_lo}).

The decomposition of data for each process is currently simply calculated by dividing the total data space by the number of
 processes used.  Therefore, for {\tt g\_lo} the blocksize for each process is calculated using a formula like this:

\begin{equation}\label{eq:totalsize}
g\_totalsize = (X*Y*l*e*s)
\end{equation}
\begin{equation}\label{eq:gblocksize}
g\_blocksize = \frac{g\_totalsize - 1}{number\ of\ processes} + 1
\end{equation}

The {\tt yxf\_lo} blocksize is calculated as follows:

\begin{equation}\label{eq:yxftotalsize}
yxf\_totalsize = (ig*isgn*inx*l*e*s)
\end{equation}
\begin{equation}\label{eq:yxfblocksize}
yxf\_blocksize = \frac{yxf\_totalsize - 1}{number\ of\ processes} + 1
\end{equation}

where {\tt inx} is the full size of {\tt X} without any points dropped to facilitate the de-aliasing filter.  Finally the {\tt xxf\_lo} blocksize is calculated as follows:

\begin{equation}\label{eq:xxftotalsize}
xxf\_totalsize = (ig*isgn*Y*l*e*s)
\end{equation}
\begin{equation}\label{eq:xxfblocksize}
xxf\_blocksize = \frac{xxf\_totalsize - 1}{number\ of\ processes} + 1
\end{equation}

Transforming data from the k-space layout {\tt g\_lo} to the layouts required for FFTs {\tt xxf\_lo} and {\tt yxf\_lo}, requires data
redistributions that distribute in the {\tt ig}, {\tt isgn} dimensions (which are local in {\tt g\_lo}), and gather in the {\tt X} and 
{\tt Y} dimensions.  If the {\tt X} and {\tt Y} data dimensions of {\tt g\_lo} are not split across processes in the parallel program, then 
these transforms simply involve moving data around in memory on each process.  However, if the {\tt X} or {\tt Y} data dimensions of {\tt g\_lo} are 
split across processes, then this swapping of data will involve sending data between processes, which is typically more costly than moving data locally.  
The number of processes used and the chosen layout can affect whether the {\tt X} and {\tt Y} data dimensions 
are split across processes or kept local to each process.

For instance, with the {\tt xyles} layout the {\tt s}, {\tt e}, and {\tt l} data dimensions of {\tt g\_lo} will be split across processes before the 
{\tt Y} and {\tt X} dimensions.  However, if the {\tt lexys} layout is used then the {\tt Y} and {\tt X} dimensions of {\tt g\_lo} will be split directly after 
{\tt s}. In the transformation from  {\tt g\_lo} to {\tt xxf\_lo}, this will require
data to be sent between processes at much smaller task counts than with the 
{\tt xyles} layout, which will impose a significant performance cost.

\section{Performance Analysis}
\label{sec:PerfAnal}
We benchmarked GS2 using a typical non-linear simulation to test the performance impact of different data layouts and process counts. 
Each simulation was run on a range of its linear sweetspot task counts.  For all
layouts, the non-linear sweetspots for {\tt xxf\_lo} and {\tt yxf\_lo} arise at 256, 512, and 1024 tasks (based on the benchmark index dimensions $l=32$, $e=8$, $s=2$, and $isgn=2$.

\subsection{Benchmarking Environment}
To analyse the performance impact of the variations in data layout on the transformations between the linear and non-linear parts 
of GS2 we ran benchmarks, using a representative simulation, on the HECToR\cite{HECToR} HPC service.  HECToR, a Cray XE6 computer, 
is the UK National Supercomputing Service.  This paper utilised the current incarnation of HECToR, Phase 3, which consists of 
nodes with two 16-core 2.3 GHz 'Interlagos' AMD Opteron processors per node, giving a total 
of 32 cores per node, with 1 GB of memory per core.  There are 928 compute nodes coupled together with 
Cray's Gemini network, providing a high bandwidth and low latency 3D torus network. The peak performance of the whole system is over 820 TF.

We used the Portland Group (PGI) Fortran compiler for the benchmarks, with the following 
compiler flags: {\bf-O3},{\bf-fastsse}.  The FFTW3 library is used to provide FFT functionality, 
and the netCDF and HDF5 libraries for I/O.  Timing information was collected using the $MPI\_Wtime()$ function, 
with each benchmark executed three  times and an average time taken.

\subsection{Benchmark Results}
\begin{figure}[t!]
\centering
\includegraphics[width=0.4\textwidth]{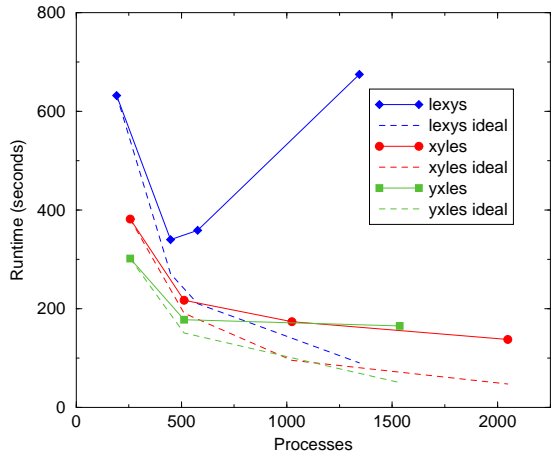}
\caption{Performance of the original code with varying process counts and layouts}
\label{fig:initperfgraph}
\end{figure}

The results collected from the benchmarks are presented in Table \ref{tab:perfanaly}.  The same results are also presented in 
Figure \ref{fig:initperfgraph} which shows the run time of each layout as a function of process count, and compares with the ideal runtimes, where the ideal run times are 
calculated with respect to the lowest process count presented 
and assuming that ideal performance is characterised by a halving in run time when then number of processes are doubled.

\begin{table}
\renewcommand{\arraystretch}{1.3}
\caption{Timings of GS2 using varying layouts and process counts}
\label{tab:perfanaly}
\centering
\begin{tabular}{||p{0.1\textwidth}|p{0.1\textwidth}|p{0.16\textwidth}||}
\hline
{\bf Layout} & {\bf Process Count} & {\bf Runtime (seconds)} \\
\hline
\multirow{4}{*}{\tt xyles} & 256 &  382 \\
\cline{2-3}
 & 512 & 217 \\
\cline{2-3}
 & 1024 & 174 \\
\cline{2-3}
 & 2048 & 138 \\
\cline{1-3}
\multirow{3}{*}{\tt yxles} & 256 & 302 \\
\cline{2-3}
 & 512 & 178 \\
\cline{2-3}
 & 1536 & 165 \\
\cline{1-3}
\multirow{4}{*}{\tt lexys} & 192 & 666 \\
\cline{2-3}
 & 448 & 371 \\
\cline{2-3}
 & 576 & 401 \\
\cline{2-3}
 & 1344 & 760 \\
\cline{1-3}
\hline
\end{tabular}
\end{table}

For the results that are presented we can see that the choice of layout can have a significant impact on the performance of GS2.  Each 
of the tests is using the same input data set, running the same simulation.  The only differences are the process counts used and the specified layout. 
We can see that in this case the {\tt lexys} \footnote{Note that the {\tt lexys} layout was developed for GS2 simulations where collision physics favours 
keeping the velocity space indices {\tt e,l} more local.} layout performs very poorly compared to the other two.  Indeed, the runtime stops decreasing with task 
count above 448 processes. (Recall that these runs were conducted with optimal process counts for the linear calculations).  On reflection it is obvious that 
{\tt lexys} is a poor choice for this calculation on these process counts, as with more than 2 processes ({\tt s} was 2 in this simulation) 
the {\tt Y} and then the {\tt X} indexes of the {\tt g\_lo} data array will be distributed across processes.  However, users would not generally select this layout for 
this type of problem and process count specifically for this reason, but it is included here to highlight the importance of the data layouts on the performance of the 
parallel program.

We can also see from Figure \ref{fig:initperfgraph} that all the layouts scale well up to around 500 processes, but that the {\tt xyles} layout 
scales significantly better than the {\tt yxles} layout at large process counts even though {\tt xyles} is using more processes (2048 compared to 1536) and therefore 
might be expected to have more parallel overheads.  The speedup for {\tt xyles} is around 2.77 times when going from 256 to 2048 processes (ideal would be 8 times), 
whereas the speedup for {\tt yxles} going from 256 to 1536 processes is around 1.82 times (ideal would be 6 times).  It should be noted that these speedups may seem to be 
low for an efficient parallel program, but at the higher process counts in these
calculations the data must be split across processes. This requires additional MPI communications to replace the less demanding local memory copies 
that can be used at lower task counts.

To investigate this performance difference further we performed some in depth profiling on {\tt xyles} at 2048 processes and {\tt yxles} on 1536 processes.  
For this profiling we collected some data (averaged across the processes running the simulation) on the MPI communications performed, which 
is presented in Table \ref{tab:profileres}.  The data on MPI message sizes is summarised by reporting statistics on groups of message sizes.

\begin{table}
\renewcommand{\arraystretch}{1.3}
\caption{Profiling results of GS2 using the {\tt xyles} and {\tt yxles} layouts}
\label{tab:profileres}
\centering
\begin{tabular}{||p{0.05\textwidth}|p{0.05\textwidth}|p{0.05\textwidth}|p{0.1\textwidth}|p{0.1\textwidth}||}
\hline
{\bf Layout} & {\bf Process Count} & {\bf Data Transform} & {\bf Profile Metric} & {\bf Profile Result} \\
\hline
\multirow{12}{*}{\tt xyles} & \multirow{12}{*}{2048} & \multirow{3}{*}{\begin{minipage}{0.05\textwidth}{\tt g\_lo} $\rightarrow$ {\tt xxf\_lo}\end{minipage}} & MPI Msg Bytes &  1499904000.0 \\
\cline{4-5}
& & & MPI Msg Count &  12000.0 msgs \\
\cline{4-5}
& & &  64KB $\le$ MsgSz $<$ 1MB Count  & 12000.0 \\
\cline{3-5}
&  & \multirow{3}{*}{\begin{minipage}{0.05\textwidth}{\tt xxf\_lo} $\rightarrow$ {\tt yxf\_lo}\end{minipage}}  & MPI Msg Bytes & 49152000.0 \\
\cline{4-5}
& & & MPI Msg Count &  4000.0 msgs \\
\cline{4-5}
& & &  4KB $\le$ MsgSz $<$ 64KB Count  & 4000.0 \\
\cline{3-5}
&  & \multirow{3}{*}{\begin{minipage}{0.05\textwidth}{\tt yxf\_lo} $\rightarrow$ {\tt xxf\_lo}\end{minipage}}  & MPI Msg Bytes & 374976000.0  \\
\cline{4-5}
& & & MPI Msg Count &  3000.0 msgs \\
\cline{4-5}
& & &  64KB $\le$ MsgSz $<$ 1MB Count  & 3000.0 \\
\cline{3-5}
&  & \multirow{3}{*}{\begin{minipage}{0.05\textwidth}{\tt xxf\_lo} $\rightarrow$ {\tt g\_lo}\end{minipage}}  & MPI Msg Bytes & 12288000.0   \\
\cline{4-5}
& & & MPI Msg Count &  3000.0 msgs \\
\cline{4-5}
& & &  4KB<= MsgSz $<$ 64KB Count  & 1000.0 \\
\cline{3-5}
\cline{1-5}
\multirow{20}{*}{\tt yxles} & \multirow{20}{*}{1536} & \multirow{5}{*}{\begin{minipage}{0.05\textwidth}{\tt g\_lo} $\rightarrow$ {\tt xxf\_lo}\end{minipage}} & MPI Msg Bytes &  2006592000.0 \\
\cline{4-5}
& & & MPI Msg Count &  12450.5 msgs \\
\cline{4-5}
& & &  256B $\le$ MsgSz $<$ 4KB Count  & 125.0 \\
\cline{4-5}
& & &  4KB $\le$ MsgSz $<$ 64KB Count  & 1895.8 \\
\cline{4-5}
& & &  64KB $\le$ MsgSz $<$ 1MB Count  & 10429.7 \\
\cline{3-5}
&  & \multirow{5}{*}{\begin{minipage}{0.05\textwidth}{\tt xxf\_lo} $\rightarrow$ {\tt yxf\_lo}\end{minipage}}  & MPI Msg Bytes & 2749824000.0 \\
\cline{4-5}
& & & MPI Msg Count &  5487.0 msgs \\
\cline{4-5}
& & &  256B $\le$ MsgSz $<$ 4KB Count  & 26.0 \\
\cline{4-5}
& & &  4KB $\le$ MsgSz $<$ 64KB Count  & 380.2 \\
\cline{4-5}
& & &  64KB $\le$ MsgSz $<$ 1MB Count  & 5080.7 \\
\cline{3-5}
&  & \multirow{4}{*}{\begin{minipage}{0.05\textwidth}{\tt yxf\_lo} $\rightarrow$ {\tt xxf\_lo}\end{minipage}}  & MPI Msg Bytes & 687456000.0   \\
\cline{4-5}
& & & MPI Msg Count &  1371.7 msgs \\
\cline{4-5}
& & &  256B $\le$ MsgSz $<$ 4KB Count  & 6.5 \\
\cline{4-5}
& & &  4KB $\le$ MsgSz $<$ 64KB Count  & 95.1 \\
\cline{4-5}
& & &  64KB $\le$ MsgSz $<$ 1MB Count  & 1270.2 \\
\cline{3-5}
&  & \multirow{5}{*}{\begin{minipage}{0.05\textwidth}{\tt xxf\_lo} $\rightarrow$ {\tt g\_lo}\end{minipage}}  & MPI Msg Bytes & 501648000.0    \\
\cline{4-5}
& & & MPI Msg Count &  3112.6 msgs \\
\cline{4-5}
& & &  256B $\le$ MsgSz $<$ 4KB Count  & 31.2 \\
\cline{4-5}
& & &  4KB $\le$ MsgSz $<$ 64KB Count  & 474.0 \\
\cline{4-5}
& & &  64KB $\le$ MsgSz $<$ 1MB Count  & 2607.4 \\
\cline{1-5}
\hline
\end{tabular}
\end{table}

From the results shown in Table \ref{tab:profileres} we can see that for the transform between {\tt g\_lo} and {\tt xxf\_lo} (and the inverse) 
both {\tt xyles} and {\tt yxles} show similar characteristics, using roughly 12000 messages for the first transform.  However, given that {\tt yxles} 
is using $75\%$ of the processes that {\tt xyles} it is surprising that it is actually sending $25\%$ more data. 

When we look at the communications required for transforming from {\tt xxf\_lo} to {\tt yxf\_lo}, however, we can see there is a very large 
difference in the amount of data sent between the two layouts.  Using {\tt yxles} generates approximately 55 times more data traffic than 
{\tt xyles} with only $25\%$ more messages.  The large difference in the amount of data that is sent using a similar number of messages, 
is explained by the fact that much larger messages are communicated in the {\tt yxles} case than in the {\tt xyles} case.

\subsection{Discussion of the performance impact of data distributions}
\label{sec:sub:impact}
Data structures with two indices are used during the 2-D FFT transformation chain from k-space into position space.  
The first index is the index along which the FFT is to be performed, and the second index is of the {\tt xxf} or {\tt yxf}-type, 
suitable for 1D FFTs in the {\tt X} or {\tt Y} directions respectively.  

Here we focus on the distribution of a data structure using an {\tt xxf}-type index.  (The discussion for a {\tt yxf}-type structure 
is essentially the same, except that the first index corresponds to  {\tt Y}).
As previously discussed, the 2nd {\tt xxf}-type index is a compound of the indices {\tt Y,ig,isgn,l,e,s}.  For all layouts described by a GS2 
layout string that ends {\tt l,e,s}, the compounded indices will be ordered as above.  In a parallel GS2 calculation, the array volume 
associated with this last index gets distributed by calculating a blocksize as outlined in Equation \ref{eq:xxfblocksize}.  
If the integer divide on the right hand side leaves any remainder, the formula rounds up the blocksize to the nearest integer to ensure that all the data is allocated.
 
The ingen tool typically guides the user to select a process count that divides the {\tt g\_lo} cleanly across cores. With 
layout {\tt xyles} or {\tt yxles}, ingen will recommend core counts where the indices  {\tt les} are well distributed for {\tt g\_lo}. 
Any recommended core count that exceeds the product {\tt les}, will be an integer multiple ${\tt j}*{\tt les}$, and will have {\tt l,e,s} maximally distributed. 
At such core counts, the  {\tt xxf\_lo} layout, with compound index ordered {\tt Y,ig,isgn,l,e,s}, will also have {\tt l,e,s} maximally distributed. 
Therefore, the remaining elements of the {\tt xxf\_lo} compound index, {\tt Y,ig,isgn}, must be distributed over {\tt j} processes.  While {\tt isgn} has 
a range of 2, the allowed range for {\tt ig} is always has an odd number which commonly does not factorise well (or may even be prime).
We will see shortly that this can lead to blocksizes for {\tt xxf\_lo} and {\tt yxf\_lo} from
Equations~(\ref{eq:xxfblocksize}) and (\ref{eq:yxfblocksize}), that have unfortunate consequences for communication.

\begin{figure}[!t]
\centering
\includegraphics[width=2.5in]{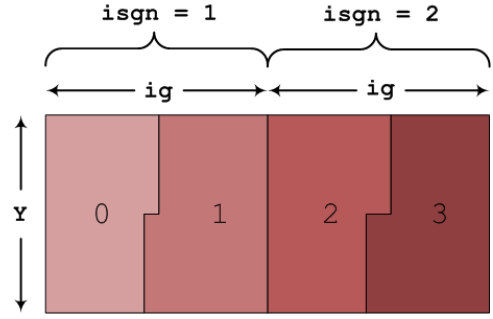}
\caption{Example of how the array projection spanned by {\tt Y, ig, isgn} will be distributed between 4 processes, when the allowed 
range for {\tt Y} is divisible by two.The colour indicates the process-id to which each piece of data is assigned.}
\label{fig:figure1}
\end{figure}

Figure \ref{fig:figure1} shows the situation where the three indices {\tt Y,ig,isgn} can be evenly divided across four processes, which is
only possible if {\tt Y} is even. As {\tt Y} is the fastest index, the split between the data going to the lower or the higher task 
rank will occur in the middle of a {\tt Y}-column.

\begin{figure}[!t]
\centering
\includegraphics[width=2.5in]{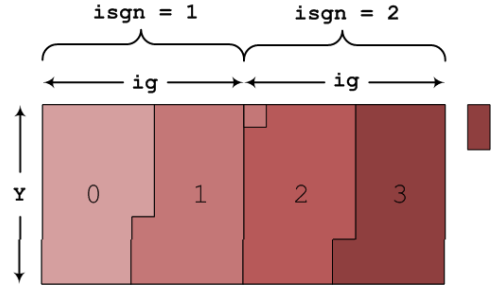}
\caption{Example of a decomposition that does not evenly split.  In this example process 0 gets more than \textonehalf \ of the first 
{\tt Y-ig} box, and process 1 is then assigned some data points from the second box.  When all the data from the first two boxes are 
assigned,  process 3 will still require data from the next pair of {\tt Y-ig} boxes, as indicated by the small box to the right of the 
main picture.}
\label{fig:figure2}
\end{figure}

If in our example, the allowed range for {\tt Y} were odd the work would not divide evenly over 4 processes, as shown in \ref{fig:figure2} 
where the sub-array spanned by {\tt Y, ig, isgn} is not exactly divisible by the number of processes and this
results in most processes being allocated the same amount of data, but can very easily result in the situation, especially for large 
problems and process counts, where one or more of the last processes in the simulation will have little or no work assigned to them (i.e. their blocksize will be small 
or zero).  While for large task counts this will not result in a significant load imbalance in the computational work, it can dramatically increase the level of 
communication that is required in the redistribution routines between {\tt xxf\_lo} and {\tt yxf\_lo}.

\begin{figure}[!t]
\centering
\includegraphics[width=2in]{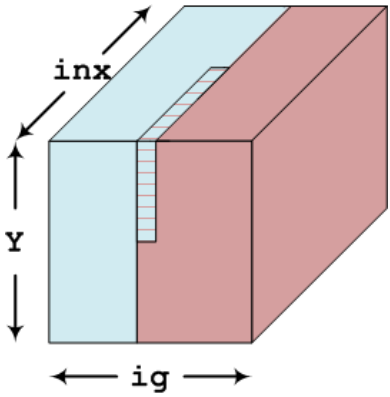}
\caption{Example of the data to be transferred between cores when transposing from {\tt xxf\_lo} to {\tt yxf\_lo}.   For {\tt xxf\_lo}, the lower 
ranking process holds the blue data and the shaded region.  The higher ranked process holds the red data.  During the transformation the 
shaded region needs to be transferred from the lower to the higher rank.  There is a similar region in the bottom rear of the cube, 
not visible in the figure, which needs transferring from the higher ranking to the lower ranking process.}
\label{fig:figure3}
\end{figure}

We now consider the amount of data that must be passed between different processes during the transpose transformation between 
{\tt xxf\_lo} and {\tt yxf\_lo}.  Figure \ref{fig:figure3} illustrates what happens when both indices split evenly 
across the processes, clarifying that only  a small amount of the data held by a process needs to be transferred between neighbouring processes.
Figure \ref{fig:figure2} showed that, if {\tt xxf\_lo} or {\tt yxf\_lo} data spaces do not divide  evenly across all processes, the equal blocksizes 
allocated to each process a will ensure all the data is allocated, but not that all processes will be allocated data.  Figure \ref{fig:figure2} demonstrates 
that this leads to an overspill, where some 
processes are allocated data from more than one {\tt l, e,} or {\tt s} index.  Furthermore, not all processes will be allocated with the full blocksize: 
there will be at least one task with a smaller allocation of data, and there may be further tasks which are allocated with {\bf no} data.    
The number of idle processes can be calculated as shown in Equation \ref{eq:xxfidleprocs} (where $xxf\_totalsize$ is defined in Equation \ref{eq:xxftotalsize}
and $xxf\_blocksize$ is defined in Equation \ref{eq:xxfblocksize}).

\begin{equation}\label{eq:xxfusedprocs}
xxf\_usedprocs = \frac{xxf\_totalsize}{xxf\_blocksize} 
\end{equation}

\begin{equation}\label{eq:xxfidleprocs}
xxf\_idleprocs = number\:of\:processes - xxf\_usedprocs
\end{equation}

Similarly the idle processes when using {\tt yxf\_lo} can be determined as follows:

\begin{equation}\label{eq:yxfusedprocs}
yxf\_usedprocs = \frac{yxf\_totalsize}{yxf\_blocksize} 
\end{equation}
\begin{equation}\label{eq:yxfidleprocs}
yxf\_idleprocs = number\:of\:processes - yxf\_usedprocs
\end{equation}

If the numbers {\tt yxf\_idleprocs} and {\tt xxf\_idleprocs} differ  significantly, it can easily be shown that large MPI messages 
will be required in the transforms between {\tt xxf\_lo} and {\tt yxf\_lo}.  Indeed this can easily arise, as the spatial dimensions in the compound
{\tt xxf\_lo} and {\tt yxf\_lo} indices, {\tt Y} and {\tt inx} respectively, are different. {\tt Y} and {\tt inx} may have quite different prime factors, 
and one layout may split evenly with no idle processes while the other does not.

\begin{figure*}[!t]
\centering
\includegraphics[width=5in]{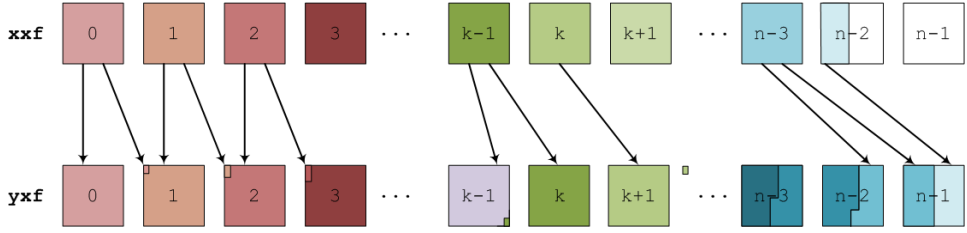}
\caption{Example of the data redistributions required when {\tt xxf\_lo} and {\tt yxf\_lo} have different numbers of idle processors. 
The colours label the data regions that are stored by each processor in {\tt xxf\_lo}.}
\label{fig:figure4}
\end{figure*}

Figure \ref{fig:figure4} demonstrates clearly how the amount of data to be sent to a different process increases linearly with increasing task 
number.  If the difference between {\tt yxf\_idleprocs} and {\tt xxf\_idleprocs} is larger than 1, the highest ranking processes (those of rank $k$ and 
above in Figure \ref{fig:figure4}) will have to transfer all of their data to different processes. Where the difference is less than one, all processes will keep some 
of their data.  

\subsection{Analysis of communication costs}
We now estimate the amount of data that must be communicated via MPI for these two cases. It is helpful to define variables

\begin{equation}\label{eq:deltaidleprocs}
delta\_idle\_proc = dabs(yxf\_idleprocs - xxf\_idleprocs)
\end{equation}
\begin{equation}\label{eq:totalredistdata}
total\_redist\_data = inx * xxf\_totalsize
\end{equation}

The total amount of data to be transferred during the redistribution (e.g. inserted into the communication network) can be estimated 
as shown in Listing \ref{alg:totaltransdata}.

\begin{lstlisting}[caption={Total amount of data to be be transferred},label={alg:totaltransdata},mathescape]
if ( delta_idle_proc .le. 1 ) then
  total_trans_data = 0.5D0 * delta_idle_proc * total_redist_data
else
  total_trans_data =  (1.0d0 - 1.0d0/(2.0d0 * delta_idle_proc)) * total_redist_data
endif
\end{lstlisting}

These estimates should be correct for large task counts, which is the situation of interest in this paper.  As the difference between 
the numbers of idle processors approaches 1, a significant portion of the data held during the transformation needs to be communicated 
between processes across the network.  This has to be contrasted with the case where the data splits evenly across processors.  In that 
case, a much smaller portion of the total data has to be passed via MPI, typically only a few percent of the total data space.

\section{Unbalanced Decomposition}
\label{sec:Unbalanced}
The functionality that transforms data from {\tt g\_lo} to {\tt xxf\_lo} involves 
moving from a data distribution when {\tt ig} and {\tt isgn} are local on each process (i.e. each process has the full dimensions of 
{\tt ig} and {\tt isgn} for a given combination of {\tt x,y,l,e,s}) to one where {\tt X} is local (i.e. each process has the full 
dimension of {\tt inx} for a given combination of {\tt Y,ig,isgn,l,e,s}). The transformation to {\tt yxf\_lo} is to a data distribution 
where {\tt iny} (where {\tt iny} full size of {\tt Y} without any points dropped to facilitate the de-aliasing filter) is guaranteed to be local 
(for any given combination of {\tt inx,ig,isgn,l,e,s}). 

When running GS2 on large process counts ($>$ {\tt l * e * s} processes) these transformations will require
unavoidable data communications, particularly the transform from {\tt g\_lo} to {\tt xxf\_lo} (which moves from {\tt ig,isgn} local to 
{\tt inx} local) as at high core counts the {\tt xxf\_lo} data distribution 
will have to split up {\tt isgn} and possibly {\tt ig} (depending on the process count) which are local dimensions in {\tt g\_lo}.

However, the complexity of the redistribution and the quantity of data to be communicated will depend on how split up the data is.  
If the data dimensions to be redistributed are only split across pairs of processes in a balanced fashion then the required message sizes and 
number of messages will be lower than if the data must be split up across a larger number of different processes.

Furthermore, if the decomposition is undertaken optimally it should be possible to ensure that the redistribution between {\tt g\_lo} 
and {\tt xxf\_lo} keeps both {\tt inx} and {\tt Y} as local as possible, thereby reducing or eliminating completely 
the MPI communication cost of the {\tt xxf\_lo} to {\tt yxf\_lo} step.

Therefore, we have created unbalanced decomposition functionality to optimise the data distribution in the {\tt xxf\_lo} and {\tt yxf\_lo}
data layouts.  This replaces the current code to calculate the block of {\tt xxf\_lo} and {\tt yxf\_lo} data that each process owns, 
moving from a uniform blocksize to adopting {\em two different blocksizes} for the process counts when the data spaces do not exactly divide by 
the number of processes used.

The new, unbalanced decomposition, uses the process count to calculates which indices can be completely 
split across the processes.  This is done by iterating through the indices in the order of the layout (so for the {\tt xxf\_lo} data 
distribution and the {\tt xyles} layout this order would be {\tt s,e,l,isgn,ig,Y}) dividing the number of processes by each index until 
a value of less than one is reached.  At this point the remaining number of processes is used, along with the index to be divided,
to configure an optimally unbalanced decomposition, by deciding on how to split the remaining indices across the remaining cores that are available.
If this index dimension is less than the number of cores available, then the index dimension is multiplied by the following index dimension until a 
satisfactory  decomposition becomes possible.

\begin{figure}[!t]
\centering
\includegraphics[width=2.5in]{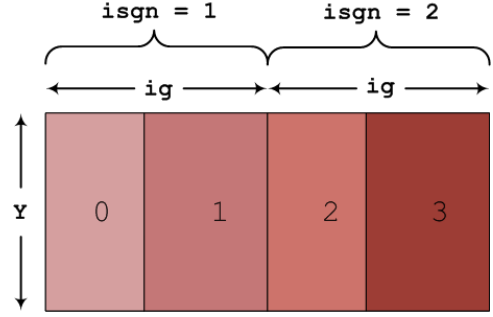}
\caption{Example of an decomposition that does not evenly split but has been decomposed using the unbalanced decomposition functionality.}
\label{fig:figure5}
\end{figure}

Figure \ref{fig:figure5} demonstrates the decomposition of the example in Figure \ref{fig:figure2} when computed using the new unbalanced 
decomposition functionality.

The drawback of using an unbalanced decomposition approach is that this makes a controlled sacrifice by introducing {\it computational} load imbalance (i.e. some processes are allocated more 
computational work than others) which is generally to be avoided for parallel programs.  The assumption that we are making is that 
the modest computational load imbalance introduced is much less significant than the drastic reduction in communications that can be achieved.  However, this assumption will only 
hold true if the amount of computational imbalance is not too large.  A computational imbalance of $100\%$ (so some processes acquire twice as much computational work as others for the 
non-linear calculations) is likely to negate the communication performance improvement from the unbalanced 
decomposition.  Therefore, with the new decomposition functionality we allow users to specify a maximum threshold to the amount of computational 
imbalance allowed.  If this threshold would be exceeded by using the unbalanced decomposition algorithm, then the original decomposition functionality 
with uniform blocksizes is utilised.

\section{Unbalanced Performance}
\label{sec:UnbalancedPerfAnal}
\begin{figure}[!t]
\centering
\includegraphics[width=0.4\textwidth]{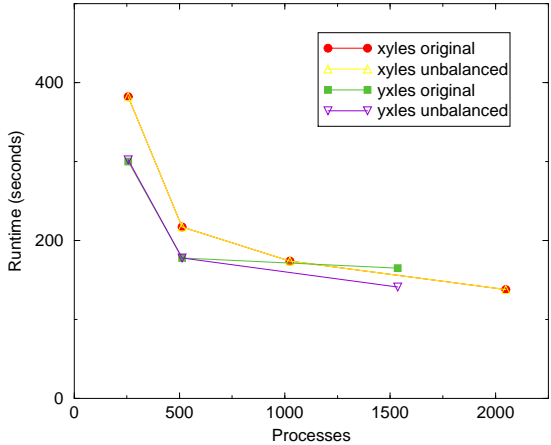}
\caption{Performance comparison of the unbalanced code with the original results collected.}
\label{fig:unbalancedperfgraph}
\end{figure}

We assessed the impact of the new unbalanced decomposition functionality on the performance of two of the simulations that appeared in Table \ref{tab:perfanaly}: the 
{\tt xyles} layout running on 2048 processes; and the {\tt yxles} layout running on 1536 processes.  The run times of the original code and of the new unbalanced code 
are compared in Table \ref{tab:unbalancedresults}, and in Figure \ref{fig:unbalancedperfgraph}.
For the {\tt yxles} layout the imbalance created by the new code (the difference in size between the small and large blocks) is approximately $5\%$ and for {\tt xyles} 
it is approximately $7\%$.

\begin{table}
\renewcommand{\arraystretch}{1.3}
\caption{Comparison of the original and unbalanced GS2 runtimes}
\label{tab:unbalancedresults}
\centering
\begin{tabular}{||p{0.05\textwidth}|p{0.05\textwidth}|p{0.1\textwidth}|p{0.07\textwidth}||}
\hline
{\bf Layout} & {\bf Process Count} & {\bf Version} & {\bf Runtime (seconds)} \\
\hline
\multirow{2}{*}{\tt xyles} & \multirow{2}{*}{2048} &  Original & 138 \\
\cline{3-4}
& & Unbalanced & 138 \\
\cline{1-4}
\multirow{2}{*}{\tt yxles} & \multirow{2}{*}{1536} & Original & 165 \\
\cline{3-4}
& & Unbalanced & 141  \\
\cline{2-3}
\cline{1-4}
\hline
\end{tabular}
\end{table}

The 1536 process simulations {\tt yxles} simulations demonstrate that the unbalanced code significantly reduces 
the overall runtime of the original code, with the unbalanced optimisation
saving around $15\%$ of the GS2 runtime.  This is in spite of our introduction 
of a computational load imbalance of around $5\%$ for the non-linear
calculations. Thus it is clear that the unbalanced decomposition can significantly improve GS2's performance.

However, if we look at the 2048 process {\tt xyles} results the unbalanced optimisation makes little difference to 
the overall runtime.  Further in depth profiling of the code reveals that the
unbalanced decomposition does significantly reduce the runtime associated with the MPI communications, 
but {\em this communications overhead is much less significant for this simulation layout and process count}.  
The MPI communications for the {\tt yxles} layout on 1536 processes takes around 6 times as long as for the {\tt xyles} layout 
on 2048 processes.  The unbalanced optimisation still does exactly what we would expect in the {\tt xyles} simulation on 2048 processes: 
it significantly reduces the cost of the communications, but it {\em does not significantly reduce the overall runtime as 
this simulation requires much less MPI communication}.

Performance profiling reveals that the unbalanced functionality reduces the requirement of MPI communications in the {\tt xxf\_lo} to {\tt yxf\_lo}
transformation in both test cases. Table \ref{tab:profileres} shows that the MPI communications required for the {\tt xyles} run were anyhow very small
(small numbers of small messages), but that MPI messages were much more significant in both size and number for the {\tt yxles} run.

This performance difference is not solely  predicated on the layout used, but also on the number of processes used.  In the benchmark
simulation using {\tt xyles} layout on 2048 processes, the {\tt xxf\_lo} total size exactly divides by the number of processes, allocating every
process with a blocksize of 496.  In a simulation with the same layout using only 1536 processes, the same data space cannot be divided
across all processes with an equal blocksize. The exact blocksize would be $661\frac{1}{3}$ in this case, so the original code rounds this up to
the nearest integer and sets a uniform
blocksize of 662. This results in the final process being allocated with no {\tt xxf\_lo} data and the second last process being allocated with only half
a block, thus generating a large communication overhead to transform the data, as illustrated in Figure \ref{fig:figure4}.

\section{Conclusions}
\label{sec:Conclusions}
We have investigated the performance of the GS2 simulation code and discovered that the choices made to incorporate both linear 
and non-linear calculation (in k-space and position space respectively) in a single code can introduce performance issues for 
a parallel code.  In particular, a process count that is optimal for the linear computation may not be optimal for the non-linear
computation (which requires redistribution of the large data arrays), and the optimal linear process count may adversely affect the performance of
the non-linear computation. Conversely, choosing a process count that is optimal for the non-linear computations, could adversely impact on the 
performance of the linear computations

Transformations between different layouts are required in order to carry out different parts of the GS2 calculation. It is desirable for the data in each layout
to be divided evenly across all processes, so that in each part of the calculation, every process has an equal blocksize of data to work
with. A process count can always be chosen so that this is achieved for one of the required layouts, but an equal distribution of data in the other
layouts cannot be guaranteed. If equal blocksizes are allocated, it is possible that for some layouts there will be a number of processes that are
allocated with no data.  Even when there are only a small number of empty processes, our performance analysis has revealed that the increased MPI
communication overhead has adverse performance implications that can be very significant.

From the computational point of view there should be negligible impact on the performance of a 1536 core simulation if 1\textonehalf \ processes are
empty, as this corresponds to only $0.1$\% of the total processes.   However, we have demonstrated that for the transforms required in GS2, this small difference in 
the number of cores required to store the data in each layout, can lead to a huge difference in the amount of data requiring to be communicated: 
this can increase from a few percent to a very significant fraction of the total data set.

Making a small change to the decomposition algorithm, to use two different blocksizes instead of uniform blocksizes, enables the data space to be split more evenly across {\em
all of the processes}. This change can drastically reduce the communications that are required for transformations between layouts.

Whilst this paper has focussed on the GS2 simulation code, the main techniques and algorithms that GS2 exploits are very widely used in scientific computing:
e.g. the different data layouts that are required for the separated linear and non-linear parts of the calculation, 
and the exploitation of FFT algorithms. The functionality we have outlined here should be widely applicable in many other scientific codes.

\section*{Acknowledgments}
Adrian Jackson was funded by the EPSRC dCSE program.  Colin Roach was funded by the RCUK Energy Programme under grant EP/I501045 and by the European 
Communities under the contract of Association between EURATOM and CCFE, while Joachim Hein was funded by the EPSRC grant EP/H00212X/1.
\newpage
Access to the HECToR supercomputer was provided through EPSRC grant EP/H002081/1.


\bibliographystyle{IEEEtran}
\bibliography{IEEEabrv,unbalancedoptimisation}

\end{document}